\documentclass[a4paper,11pt]{article}
\usepackage{jinstpub}
\usepackage{lineno}
\usepackage{graphicx}
\usepackage{amsmath}
\usepackage{url}
\usepackage{booktabs}
\usepackage{array}
\usepackage{multirow}
\usepackage{threeparttable}
\usepackage{subfigure,dcolumn}
\usepackage{epstopdf}
\usepackage{mhchem}
\usepackage{upgreek}
\usepackage{graphicx}
\usepackage[section]{placeins}
\usepackage{float}
\usepackage{enumerate}

\title{\boldmath Convolutional Neural Network-Based Neutron and Gamma Discrimination in EJ-276 for Low-Energy Detection}

\author{Fengzhao Shen, Tao Li, Jingkui He, Shenghui Xie, Yuehuan Wei, Tuchen Huang, and Wei Wang}
\affiliation{Sino-French Institute of Nuclear Engineering and Technology,
Sun Yat-sen University, Tang Jia Wan, Zhuhai, China}

\emailAdd{litao73@mail.sysu.edu.cn; weiyh29@mail.sysu.edu.cn; huangtuchen@mail.sysu.edu.cn}

\abstract{Organic scintillators are important in advancing nuclear detection and particle physics experiments. Achieving a high signal-to-noise ratio necessitates efficient pulse shape discrimination techniques to accurately distinguish between neutrons, gamma rays, and other particles within scintillator detectors. Although traditional charge comparison methods perform adequately for $\sim$MeVee particles, their efficacy is significantly reduced in the lower energy region(\textless\,200\,keVee).
This paper introduces a particle identification method that harnesses the power of a convolutional neural network. We focused on the convolutional neural network's exceptional ability to discriminate between neutrons and gamma rays in the low-energy spectrum, utilizing a setup comprising a plastic scintillator EJ-276 and Silicon photomultiplier readout. Our findings reveal remarkable accuracies of 97.3\% and 98.6\% in the 0$\sim$100\,keVee and 100$\sim$200\,keVee energy ranges, respectively. These results represent substantial improvements of 13.8\% and 4.25\% over conventional methods.
The enhanced discrimination power of the convolutional neural network method opens new frontiers for the application of organic scintillation detectors in low-energy rare event searches, including dark matter and neutrino detection.}


\begin{document}
\maketitle
\flushbottom

\section{Introduction}
Organic scintillators have been active as detectors in various fields of nuclear technology and particle detection. These applications encompass critical areas such as nuclear power plants and fuel cycles\,\cite{bib:S.A. Pozzi,bib:V.L. Romodanov,bib:R.T. Kouzes,bib:M. Jalali}, fusion plasma diagnostics\,\cite{bib:X. Yuan}, nondestructive testing\,\cite{bib:J.A. Hashem}, homeland security preventing the illicit transportation of nuclear materials\,\cite{bib:R.C. Runkle,bib:H. Al Hamrashdi}, and particle physics experiments\,\cite{bib:P.K. Netrakanti,bib:J. Ashenfelter,bib:C. Stewart,bib:K.Y. Jung}. Organic scintillators are sensitive to various types of particles.
Therefore, particle identification based on scintillators is crucial to reduce the backgrounds and improve the signal-to-noise ratio (SNR). To tackle this problem, the pulse shape discrimination~(PSD) method has been intensely developed, including the charge comparison method~(CCM)\,\cite{bib:F. Ferrulli}, zero-crossing method\,\cite{bib:D. Wolski}, wavelet transform method\,\cite{bib:H. Arahmane}, and pulse-gradient analysis method\,\cite{bib:M.D. Aspinall}. The CCM is one of the most common PSD techniques. It is well-known for robust performance under diverse conditions, requiring minimal time and computational resources\,\cite{bib:F. Ferrulli,bib:R. M. Preston,bib:F. Shen2022}. However, discriminating between neutrons and $\gamma$ rays in the lower equivalent electron luminescence energy range is still challenging, especially below 200~keVee.

In homeland security, the neutron spectrum from spontaneous fission events (like $^{252}$Cf and $^{240}$Pu) and neutron-induced fission reactions ($^{235}$U) encompasses a wide energy range, wherein the low-energy part (below 200\,keV) consitutes a significant proportion\,\cite{bib:S.A. Pozzi,bib:R.C. Runkle}. Additionally, some environmental gamma rays also fall within this low-energy range. Inaccurately distinguishing between neutrons and gamma rays in this range can lead to false alarms or missed detections.
In addition, the particle physics experiments also require particle identification in the low-energy region. For example, the organic scintillator was proposed as a dark matter detection medium utilizing carbon and hydrogen nuclear recoils twenty years ago\,\cite{bib:J. Hong}. This approach has become more attractive with the development of large neutrino detectors using organic liquid scintillators, such as Borexino, SNO+, and JUNO\,\cite{bib:J. Bramante,bib:T.A. Laplace}.
The organic scintillator is also used in the Borexino solar neutrino experiment. The designed energy threshold is 50\,keV, but the analysis threshold is a few hundred keV, limited by the knowledge of background\,\cite{bib:D. D’Angelo}. This indicates the importance of particle discrimination in the low-energy region for background reduction.
The PSD technique of $n$/$\gamma$ can further reduce the energy threshold in the veto detector, thereby increasing veto efficiency and reducing background in the main detector\,\cite{bib:D.S. Akerib}.

In recent years, the ascent and progression of artificial intelligence have led to an increased application of artificial neural network methods in $n$/$\gamma$ discrimination. Some studies have shown that artificial intelligence methods can improve the anti-noise performance of $n$/$\gamma$ signal discrimination\,\cite{bib:H. Liu}, especially under high counting rates and stacking conditions\,\cite{bib:J. Han}. Some researchers use multilayer perceptron (MLP), convolutional neural network (CNN) and other artificial neural network (ANN) to highlight the advantages of low-energy discrimination\,\cite{bib:A. Hachem,bib:S.Y. Zhang,bib:M. Yoshino,bib:V. H. Hai} at a minimum energy of 70\,keVee.

In this work, a PSD method based on CNN to enhance the low-energy discrimination power has been developed.
A 1-inch plastic scintillator EJ-276 coupled with a Silicon photomultiplier (SiPM) array is utilized to investigate the PSD performance by comparing the discrimination accuracy of the traditional CCM with that of a CNN. Two CNN methods are employed, including a network model designed for time series data and another based on image analysis. The applicable energy region can be as low as 40\,keVee, which is limited by the energy threshold of the experimental setup caused by the electronic noise level, instead of the CNN algorithm.

\section{Experimental setup}
\begin{figure}[!htb]
    \centering
    \includegraphics[width=.65\hsize]{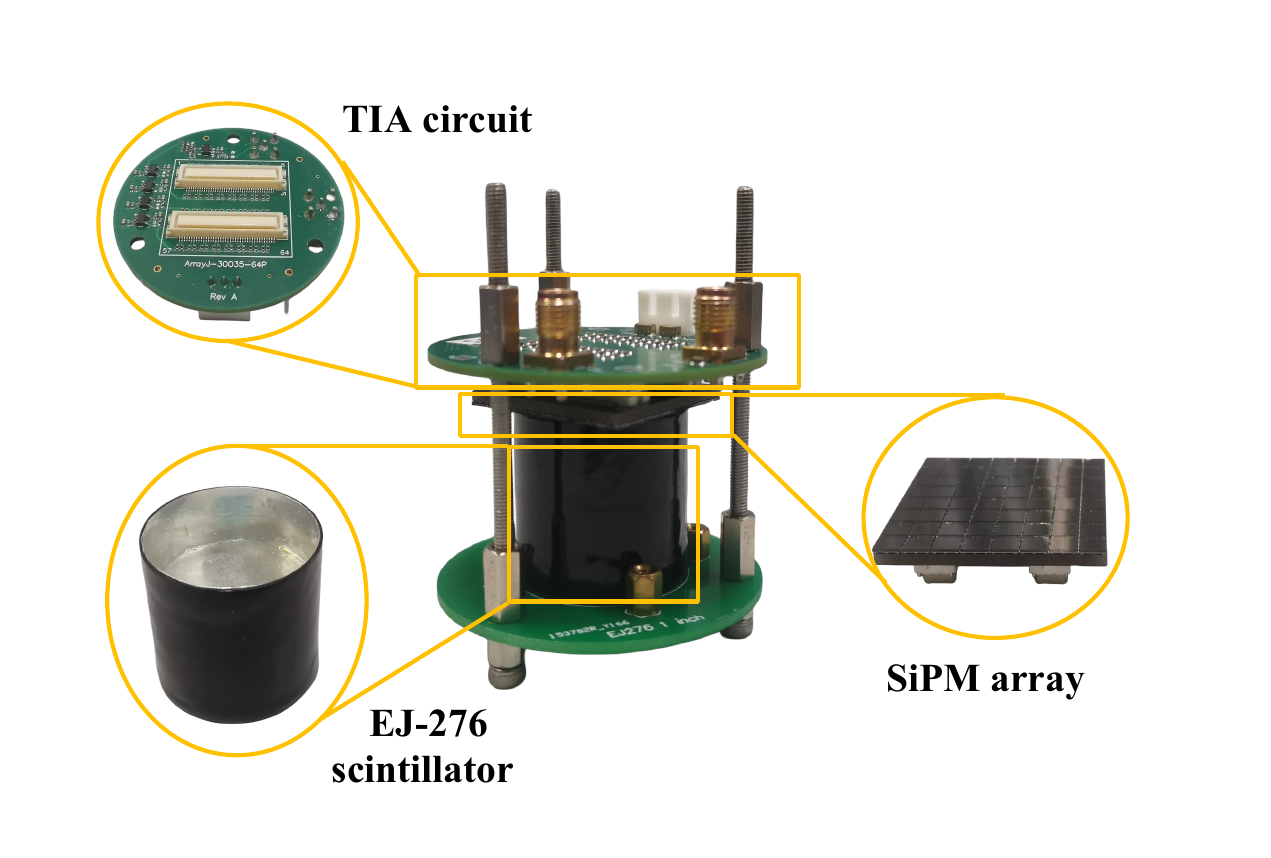}
    \caption{Detector System: Integration of EJ-276 scintillator with SiPM array.}
    \label{fig:Figure-of-assembly}
\end{figure}

Our experimental setup included an EJ-276 plastic scintillator from Eljen Technology, a cylindrical specimen measuring $\phi$ 1 × 1 inch. The silicon photomultiplier, obtained from ON Semiconductor, consisted of an 8 × 8 array, with each element containing 3 × 3 $mm^{2}$ J-series SiPM elements, as shown in Fig.~\ref{fig:Figure-of-assembly}. The SiPM array was subdivided into four groups for readout, each housing 16 SiPMs. In each group, the 16 SiPMs were connected in parallel, and their signals were amplified using a transimpedance amplifier circuit (TIA). This TIA featured a specialized transimpedance amplifier tailored for SiPM array readout, developed from our previous research\,\cite{bib:T. Huang}. The outputs from the four TIAs were combined into a single output. The bias voltage for the SiPM was set at 27.5\,V.
For data acquisition, a Pico 5444B digitizer sampled the pulse waveforms at 500\,MSPS with a 12-bit resolution. To avoid noise, the electronic threshold is set at 0.15\,V, corresponding to a minimum detection energy of 40\,keVee according to the energy calibration. The sampling serial port contains 500 points, with 200 points before triggering and 300 points after triggering. The data were then transferred to a computer for detailed offline analysis.

\begin{figure}[!htb]
    \centering
    \includegraphics[width=.65\hsize]{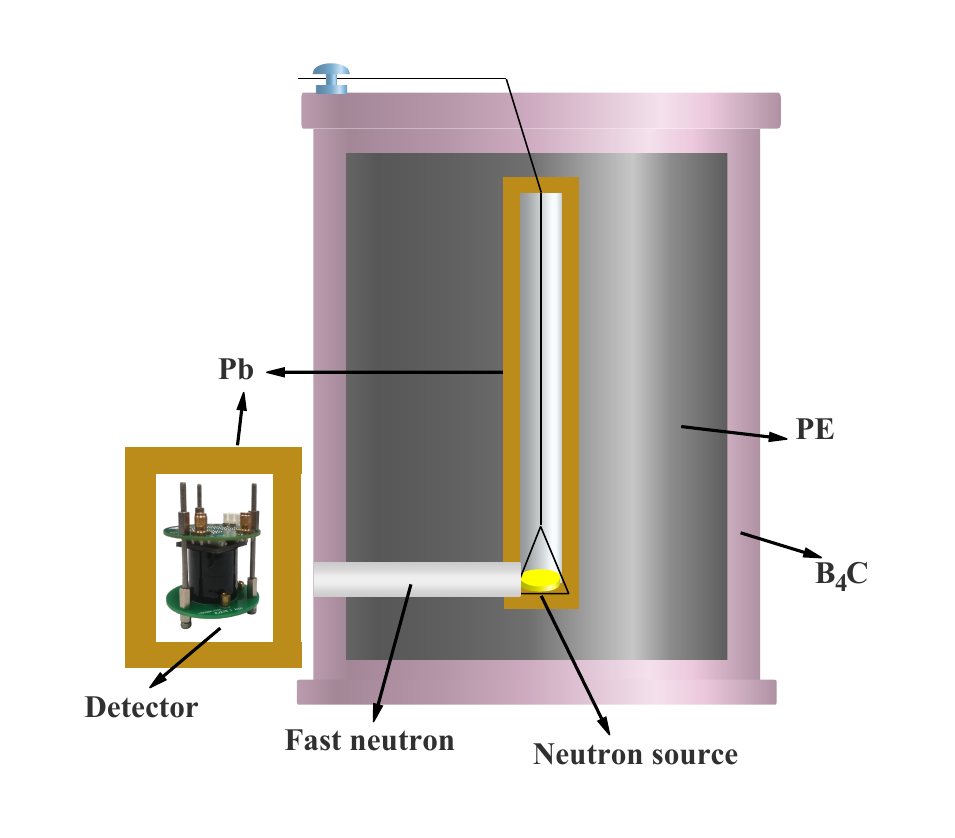}
    \caption{Neutron signal acquisition setup using an AmBe source. The neutron source container includes a shielding design with boron carbide (B$_{4}$C), plumbum (Pb) and polyethylene (PE) to ensure the dose around the container remains at background levels.}
    \label{fig:neutron source}
\end{figure}

For energy calibration, we employed gamma sources, including $^{137}$Cs, $^{60}$Co, and $^{22}$Na. Calibration involved using Compton edge energies, specifically 341\,keV and 1062\,keV for the 511\,keV and 1275\,keV gamma rays from $^{22}$Na, and 447\,keV for the 662\,keV gamma rays from $^{137}$Cs. However, distinguishing between the two Compton edges of $^{60}$Co was challenging within the spectrum. Consequently, we used the average energy of these two Compton edges, 1041\,keV, as the calibration point\,\cite{bib:E.V. Pagano}. Linear fitting of the energy scale has been performed in our previously published articles\,\cite{bib:F. Shen2022}.
For neutron signal acquisition, an Americium Beryllium neutron source ($^{241}$AmBe) with an activity of 4.81×$10^{8}$\,Bq was used. The neutron source is encapsulated in a cylindrical metal structure ($\phi$ 16×19\,mm) and housed in a specially designed container approximately 49.4\,cm in diameter and 76\,cm in height. A fast neutron aperture is located at the base of the container, with the detector positioned at its exit, as illustrated in Fig.~\ref{fig:neutron source}.

\section{Charge comparison method}
\begin{figure}
    \centering
    \includegraphics[width=.6\hsize]{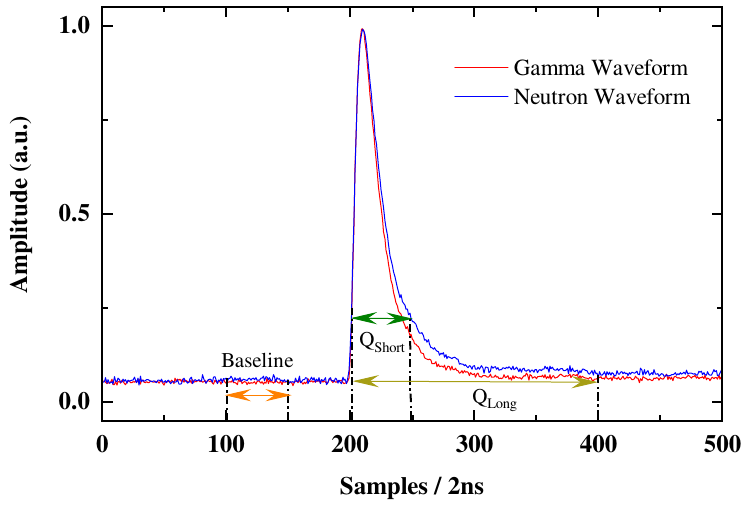}
    \caption{Measurement of gamma rays and neutrons pulse shapes and time windows using a 1-inch EJ-276 SiPM array. Points from the 100th to the 150th were averaged to determine the baseline value, which was calculated as 28.4 mV.}
    \label{fig:wave}
\end{figure}

\begin{figure}
    \centering
    \includegraphics[width=.65\hsize]{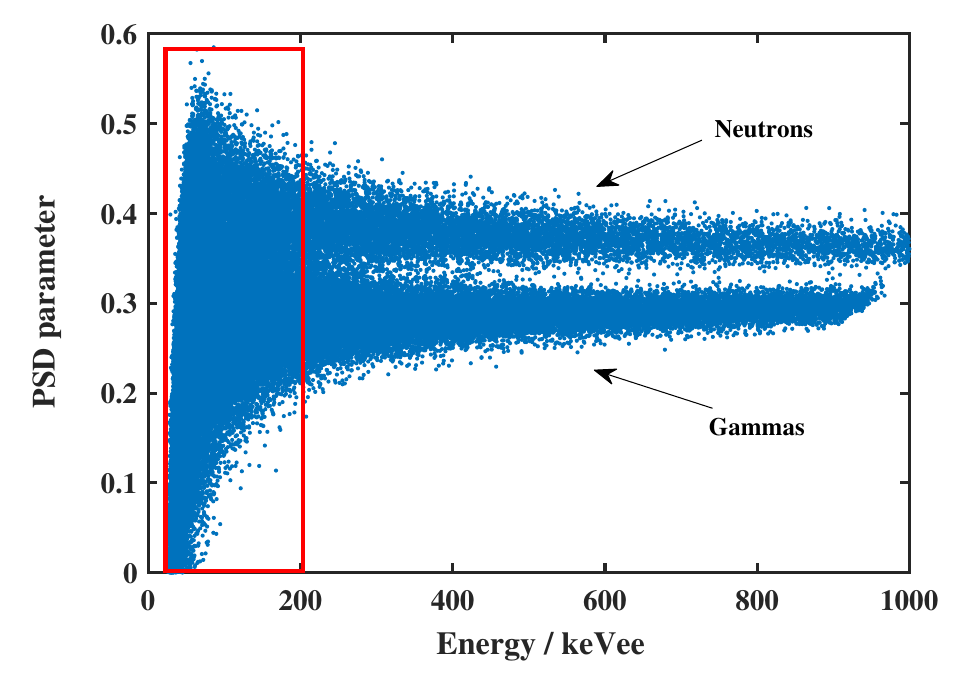}
    \caption{PSD distribution as a function of reconstructed energy from AmBe calibration. The low-energy part (\textless\,200\,keVee) can not be separated from CCM (indicated in the red box).}
    \label{fig:PSD-2D}
\end{figure}

\begin{figure}
    \centering
    \includegraphics[width=0.7\hsize]{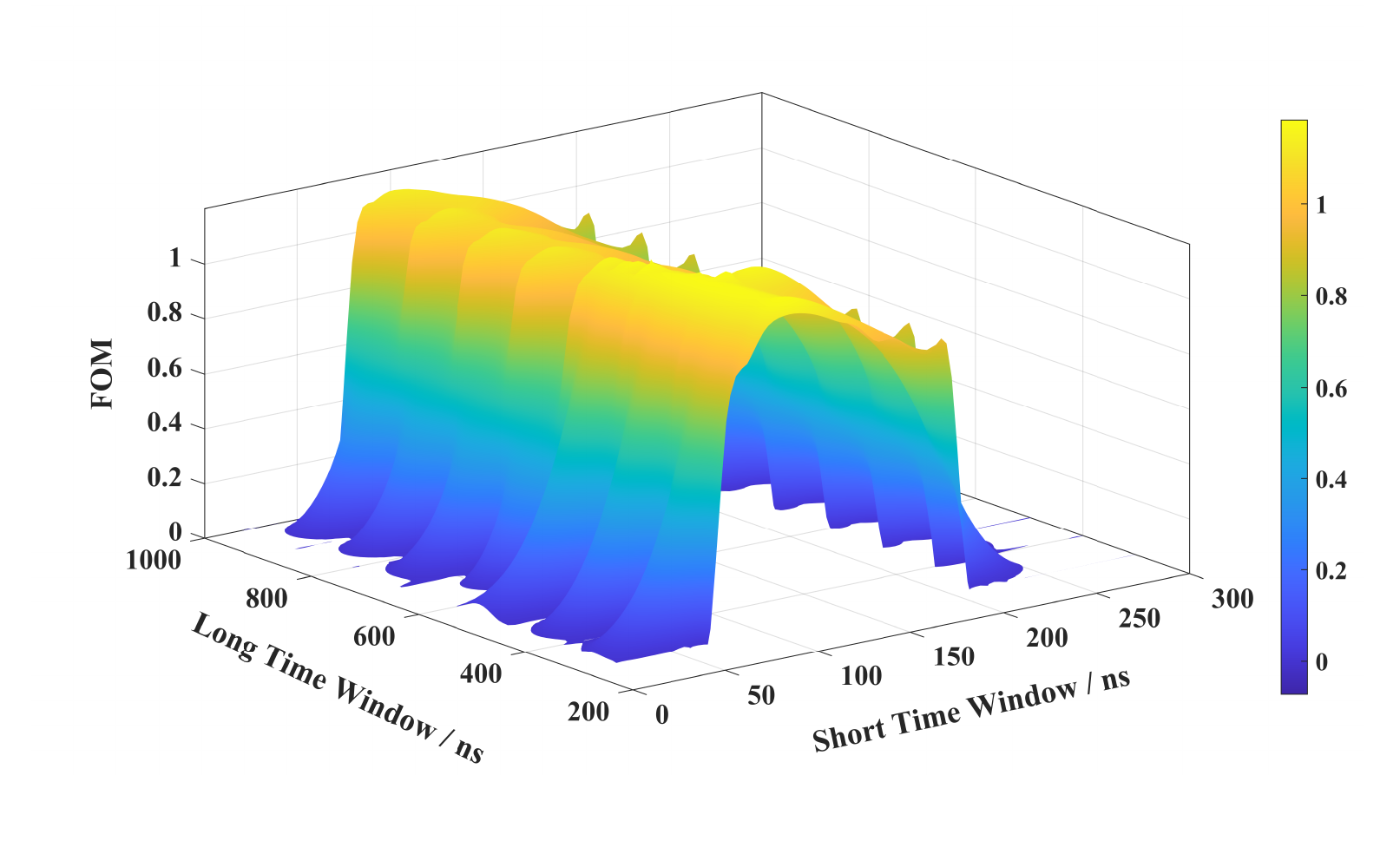}
    \caption{FoM analysis along the Z-Axis for different time window variations.}
    \label{fig:timeParameters}
\end{figure}

The conventional CCM was implemented by comparing integrated charges over two windows of different lengths.
Both neutron and gamma-induced pulses in organic scintillation have short decay (prompt) and long decay (delayed) fluorescence components. The shape of the pulse varies due to the distinct proportions of prompt and delay components exhibited by different particles.
The normalized pulse shapes resulting from neutrons and gamma rays interactions in the EJ-276 scintillator are shown in Fig.~\ref{fig:wave}. The corresponding PSD value can be calculated as follows:
\begin{equation}
\label{eq:PSD}
\begin{aligned}
PSD=\frac{Q_{Long}-Q_{Short}}{Q_{Long}}
\end{aligned}
\end{equation}
where $Q_{Short}$ and $Q_{Long}$ represent charge integrations over short and long windows, respectively, together with the baseline corrections. The performance of PSD is assessed using the FoM, defined as:
\begin{equation}
\label{eq:FoM}
\begin{aligned}
FoM=\frac{|P_{n}-P_{\gamma}|}{W_{n}+W_{\gamma}}
\end{aligned}
\end{equation}
where $P_{n}$ and $P_{\gamma}$ denote the positions of the neutrons and gamma rays peaks in the PSD distribution, fitted with Gaussian distributions, while $W_{n}$ and $W_{\gamma}$ are the full widths at half maximum (FWHM) of these peaks.
A higher FoM value indicates a clearer discrimination between neutrons and gamma rays signals.
However, variations in Compton electron interactions, pile-up effects, and noise can reduce waveform distinction at higher doses and lower thresholds when using the CCM method. This scenario reflects the conditions in the inadequate PSD region shown in Fig.~\ref{fig:PSD-2D}. Meanwhile, the choice of different integrating windows can impact the FoM value. By scanning different integrating windows, a three-dimensional distribution map can be constructed to illustrate how FoM values vary with changes in the long and short windows. As shown in Fig.~\ref{fig:timeParameters}, the optimal long and short window sizes are identified as 430\,ns and 106\,ns, respectively.

\section{Convolutional neural network}
Focus on the problem at low-energy region shown in Fig.~\ref{fig:PSD-2D}, two kinds of CNNs have been utilized to enhance the discrimination power. The CNN for the time series classification (TSC) involves two main approaches. The first approach adapts traditional CNN architectures to handle one-dimensional time series signals. The second approach transforms these signals into two-dimensional matrices, similar to traditional image recognition applications. This study utilizes and compares these two methods.

\subsection{Data preparation}

The training dataset for the classification task comprises two types of events.
Events from the $\gamma$ radioactive source are labelled as 0 while $n$ events as 1. A dataset containing 40,000 waveforms was used for model training, with an 80\% to 20\% split between training and validation sets, respectively. Additionally, a separate set of 80,000 waveforms containing a mix of neutrons and gamma rays events was reserved for model testing.

To reduce data volume and speed up training time, the original 500 sampling points were reduced to 400. The reduction was made by removing the 0th to 100th sampling points in the waveform baseline, which does not impact the pulse shape discrimination. Each waveform's event window comprises 400 sampling points, resulting in a one-dimensional time series input of size 400x1 for the model.
Waveform normalization is achieved by scaling the maximum amplitude to 1.
This normalization makes the model's training energy-independent, allowing it to focus on the waveform's structural features. Additionally, it helps prevent gradient explosion or disappearance.

\subsection{One-dimensional time series method}
\begin{figure*}[h]
    \centering
    \includegraphics[width=1.1\hsize]{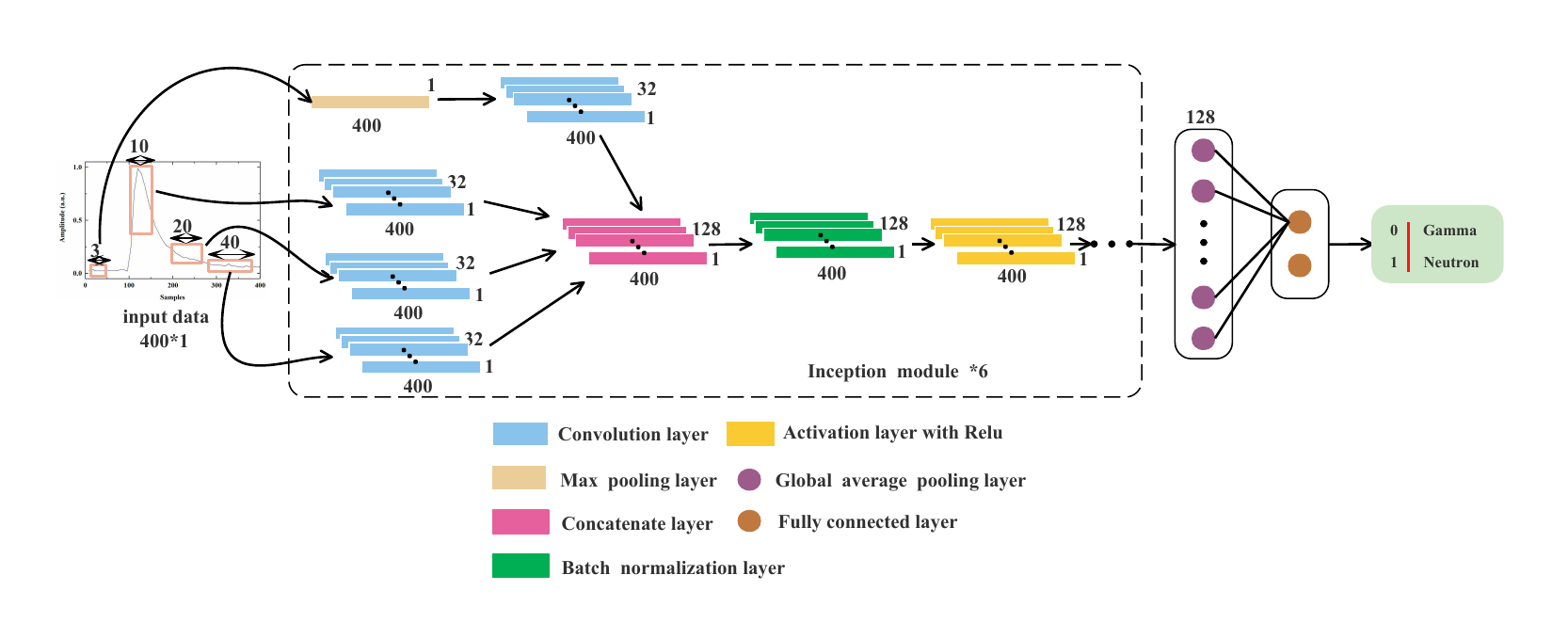}
    \caption{Structure of the InceptionTime model for one-dimensional time series method. The model incorporates six Inception modules, connected by residual links. Each Inception module utilizes three sets of filters, each containing 32 filters with lengths of l $\in$ \{10, 20, 40\} with MaxPooling added to the mix. MaxPooling applies a 3x1 convolutional kernel to a 400x1 data array, followed by the use of 32 filters—identical to those in the other three parallel processes—for feature fusion. Consequently, each Inception module has 128 filters after feature fusion. The final layer employs a sigmoid function as the activation function for the fully connected layer, mapping the variables to the range [\,0, 1\,].}
    \label{fig:InceptionTimeModel}
\end{figure*}

An InceptionTimeNet\,\cite{bib:H. Ismail Fawaz} model was employed to enhance waveform feature extraction to distinguish between neutrons and gamma rays events.
``InceptionTime", a deep learning network designed specifically for time series data, was proposed by Ismail Fawaz et al. in 2019.
Derived from Google's ``Inception" framework, this network architecture is specifically tailored for time series data.
The structure of InceptionTimeNet is shown in Fig.~\ref{fig:InceptionTimeModel}.
A key characteristic of InceptionTime is its use of convolutional kernels of varying sizes within the same layer to capture features at multiple time scales.
This feature allows InceptionTime to capture long-term and short-term data patterns effectively.
Key features of the InceptionTime network architecture include:
\begin{enumerate}[a)]
    \item Parallel convolutional layers: InceptionTime features multiple parallel convolutional layers, each using different-sized kernels to capture features at diverse time scales simultaneously.
    \item Residual connections: InceptionTime includes residual connections that alleviate the problem of vanishing gradients, facilitating easier training and more effective deep representation learning.
    \item Global average pooling: The final layer of the network employs global average pooling to transform the processed time series data into a fixed-length vector, used for downstream tasks like classification or regression.
\end{enumerate}

\subsection{Two-dimensional matrices method}

Given CNN's notable success in image classification, this study converts one-dimensional time series data into two-dimensional images. Subsequently, it redefines the time series classification task as an image classification challenge, utilizing CNN models for enhanced training and learning. To facilitate this transformation, the study employs the technique of recurrence plot (RP). This method is widely used for data processing and feature extraction, especially in analyzing time series data and pattern recognition tasks.

The RP method analyzes complex systems by transforming a time series into a two-dimensional representation, visualizing the recurrence of the system's states over time. This method excels in dealing with non-linear and non-stationary data\,\cite{bib:N. Hatami,bib:X. Li}.

To construct a recurrence plot, a time series $X = {x_1, x_2, ..., x_n}$ is first embedded into a higher-dimensional space, reconstructing its phase space trajectory using the delay embedding method. In this work, an embedding dimension of 2 is chosen to reduce simulation complexity, prevent overfitting, and minimize computational demands. The time delay is set to 1 to effectively capture sequence details, making it particularly suitable for small signals.
The standard recurrence of a state at time $i$ to a state at time $j$ is then defined by the Heaviside step function, represented mathematically as:

\begin{equation}
\label{eq:recurrence}
R_{i,j} = \Theta(\epsilon - ||\vec{x}_i - \vec{x}_j||), \quad i,j = 1, ..., N
\end{equation}

Here, $\vec{x}_i$ and $\vec{x}_j$ represent the phase space vectors at times $i$ and $j$, respectively. $\epsilon$ denotes a threshold distance, $||\cdot||$ represents a norm, and $\Theta(\cdot)$ is the Heaviside step function, which is $1$ when its argument is positive and $0$ otherwise.
The resulting matrix, $R$, is a binary matrix encoding the recurrences. When plotted, this matrix visually represents the dynamical system's behavior over time. In the matrix, points $(i, j)$ where $R_{i,j} = 1$ indicate that the states at times $i$ and $j$ are within the threshold distance $\epsilon$, signifying a recurrence of the system's states.

To effectively capture the intricate temporal dynamics and nonlinear dependencies inherent in the detector signals, we employed a modified RP transformation method. Unlike the standard RP approach, which typically results in a binary matrix indicating the recurrence of states based on a predefined threshold, our methodology preserves the continuous nature of state similarities by normalizing the RP matrix to a range of [-1,1].
First, the raw signal x was normalized to the range [0,1]:
\begin{equation}
\label{eq:recurrence-1}
X_{\text{norm}} = \frac{X - \min(X)}{\max(X) - \min(X)}
\end{equation}
Next, we reconstructed the phase space by creating state vectors using time-delay embedding with an embedding dimension of 2:
\begin{equation}
\label{eq:recurrence-2}
S_i = \left[X_{\text{norm}}(i), X_{\text{norm}}(i+1)\right]
\end{equation}
The RP matrix R was then computed by calculating the squared Euclidean distance between each pair of state vectors:
\begin{equation}
\label{eq:recurrence-3}
R(i, j) = \sum_{k=1}^{2} \left(S_i(k) - S_j(k)\right)^2
\end{equation}
Finally, the distance matrix R was normalized to the range [-1,1] using linear scaling:
\begin{equation}
\label{eq:recurrence-4}
R_{\text{normalized}} = 2 \times \frac{R - \min(R)}{\max(R) - \min(R)} - 1
\end{equation}

\begin{figure}
\centering
\subfigure[2D recurrence plot images of gamma pulses.]{
  \includegraphics[width=0.35\textwidth]{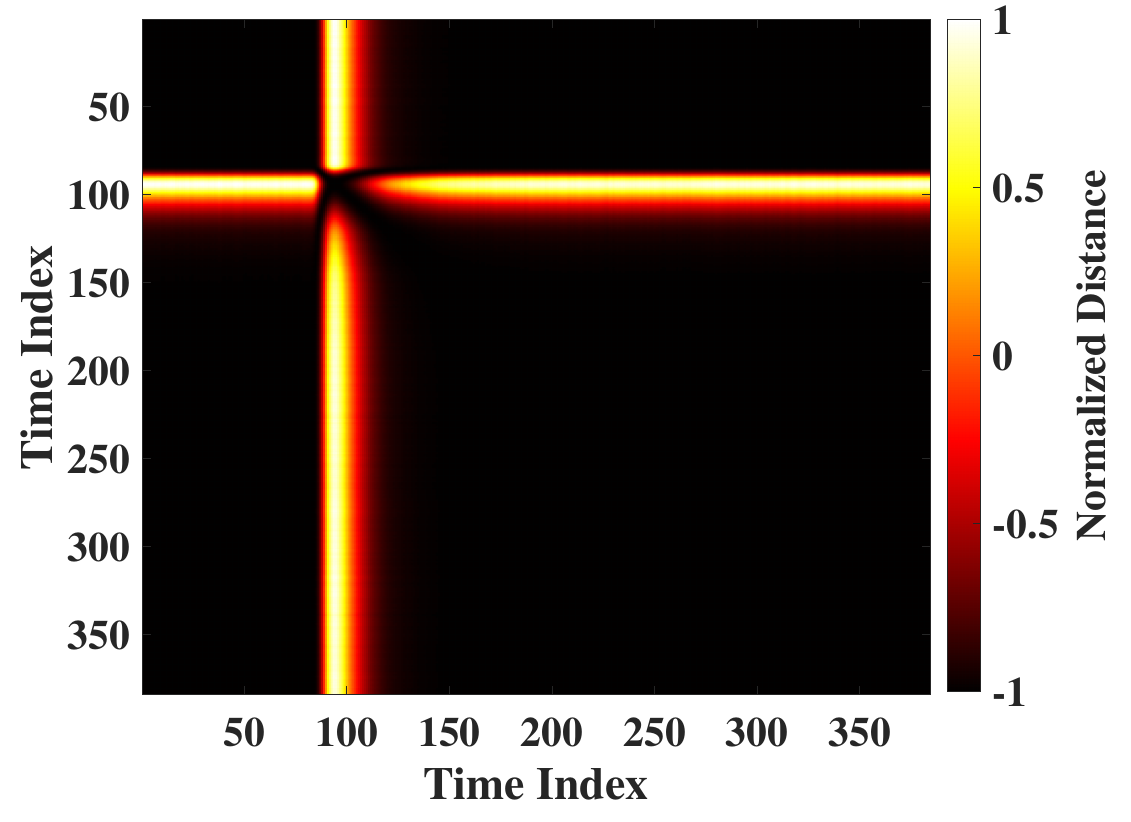}
  \label{fig:gamma-RP}
}
\subfigure[2D recurrence plot images of neutron pulses.]{
  \includegraphics[width=0.35\textwidth]{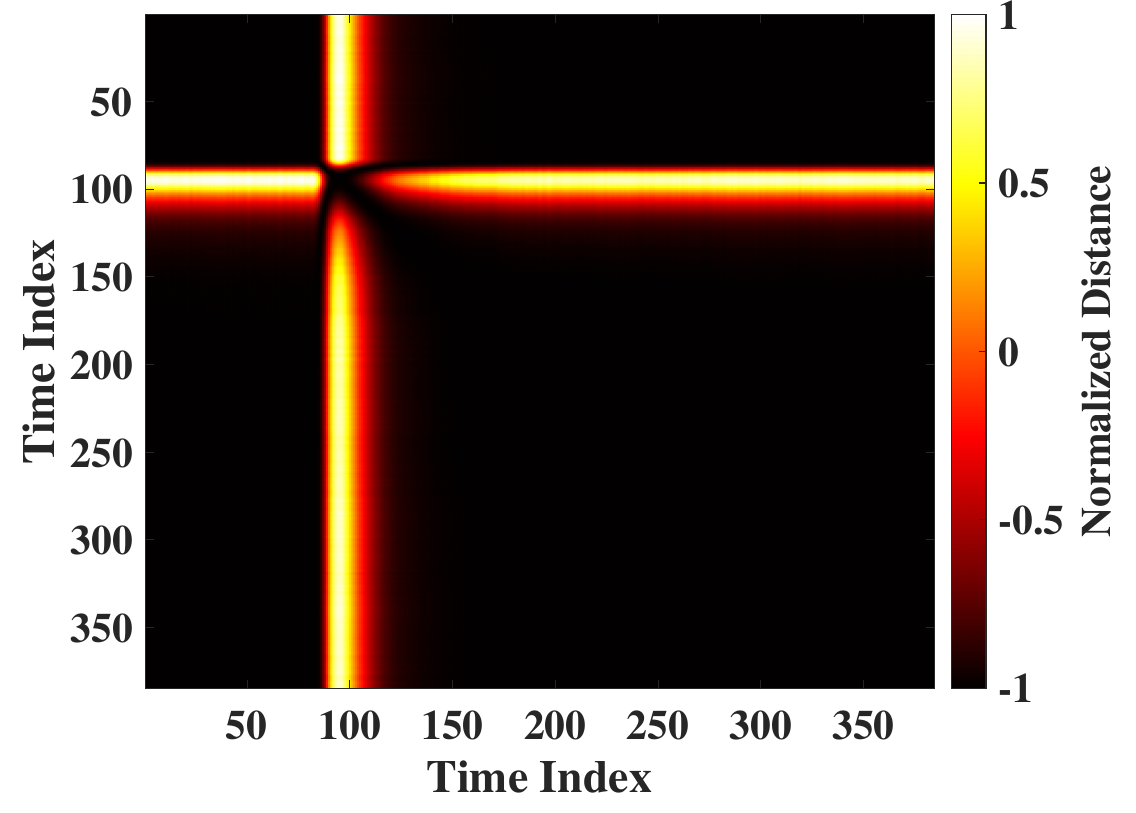}
  \label{fig:neutron-RP}
}
\subfigure[2D recurrence plot difference image of neutron and gamma, calculated as the absolute value of Fig.~\ref{fig:gamma-RP} minus Fig.~\ref{fig:neutron-RP}.]{
  \includegraphics[width=0.35\textwidth]{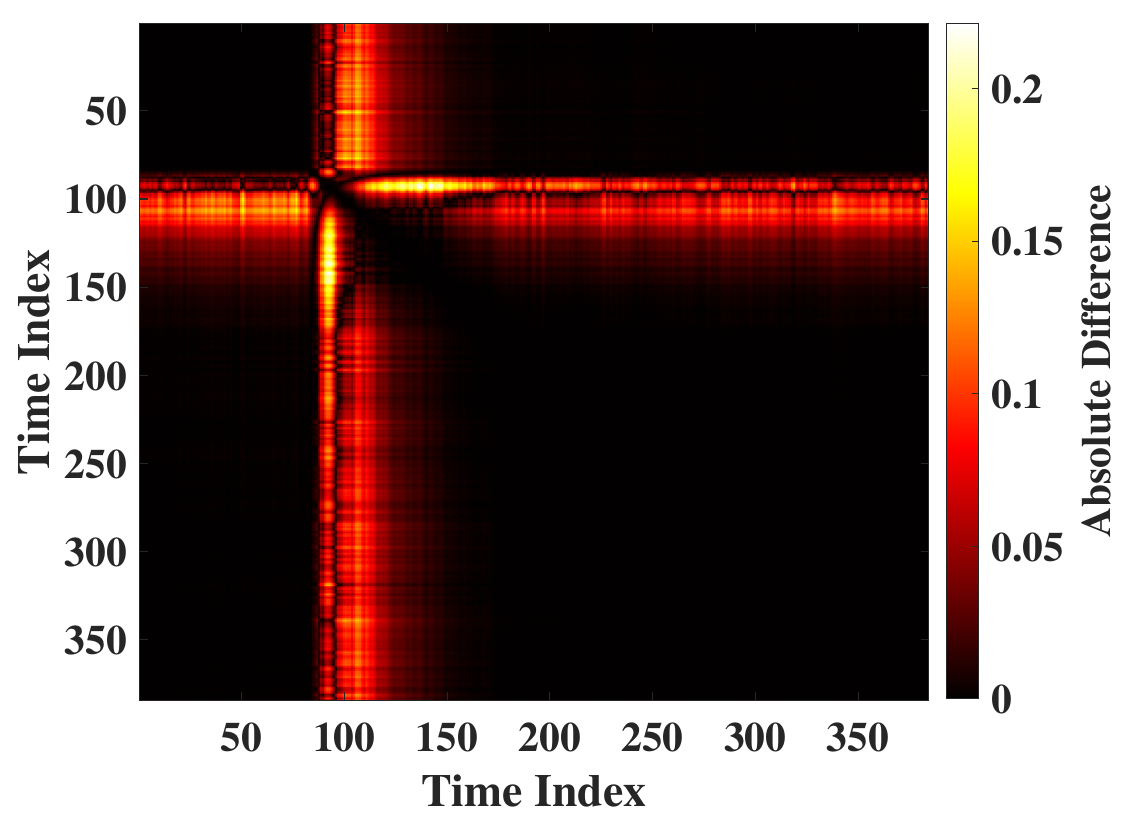}
  \label{fig:differ-RP}
}
\caption{2D recurrence plot images of gamma and neutron pulses.}
\label{fig:2D Recurrence Plot Images}
\end{figure}

This adjustment enables the retention of nuanced similarity information between different states, thereby providing a richer feature set for subsequent analysis.
The images of gamma and neutron pulses after the RP transformation are shown in Fig.~\ref{fig:gamma-RP} and Fig.~\ref{fig:neutron-RP}, respectively. Fig.~\ref{fig:differ-RP} illustrates their differences.

\begin{figure}
    \centering
    \includegraphics[width=1.05\hsize]{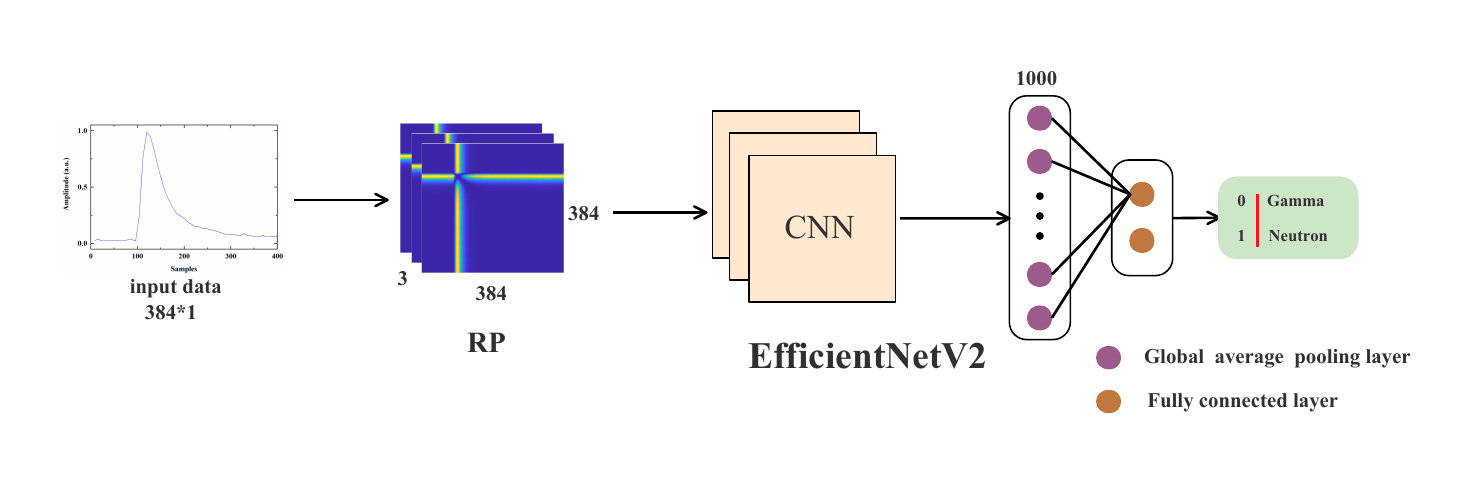}
    \caption{The two-dimensional matrices method of the EfficientNetV2 model architecture.}
    \label{fig:EfficientNetV2Model}
\end{figure}

The comprehensive architecture of this model is illustrated in Fig.~\ref{fig:EfficientNetV2Model}.
Our machine learning model utilizes the EfficientNetV2 network\,\cite{bib:M. Tan}. EfficientNetV2 introduces Fused-MBConv layers, which combine standard convolutions with batch normalization and include depthwise convolutions to optimize model efficiency and reduce the model size. It incorporates progressive learning techniques that incrementally improve the network’s resolution, thus shortening training durations. Additionally, the training process has been optimized with advanced data augmentation and regularization techniques, markedly enhancing training speed and model efficiency without compromising accuracy. These advancements provide practical solutions for implementing efficient deep learning models in resource-limited settings like embedded systems.

In this study, RP images are resized to 384x384 (depend on the CNN model) and fed into EfficientNetV2 model. To enhance training accuracy using the model's pre-trained weights, three copies of the RP image are converted into an input format with three channels. The pre-trained weights provided by the model's developer were used. The developer initially trained the model on the larger ImageNet-21K dataset and then fine-tuned it on the ImageNet ILSVRC2012 dataset to obtain these weights. In this work, the entire model was fine-tuned using the pre-trained weights to fully leverage the feature information of the new dataset and enhance the model's performance. For data augmentation and regularization, Dropout was employed as mentioned in the original paper. Specifically, dropout was applied in the EfficientNetV2-M model with a rate set to 0.3. Additionally, standard data augmentation techniques, such as random cropping and flipping, were also utilized.

\subsection{Model performance}
When tuning model hyperparameters, the focus is on batch size, learning rate and optimizer. Initially, the value of each parameter is selected, followed by fixing two parameters while adjusting the third one sequentially. Different learning rates (ranging from 1e-6 to 1e-3) are tested while keeping the batch size and optimizer constant. Next, with the optimal learning rate and optimizer fixed, the batch size is adjusted (from 8 to 256). Finally, the optimal learning rate and batch size are fixed, and different optimizers (SGD and Adam) are tested. Accuracy and loss are recorded for each adjustment to determine the optimal combination of parameters, thereby improving model performance.

\begin{figure}
\centering
\subfigure[Training loss and accuracy of the InceptionTime model.]{
  \includegraphics[width=0.6\textwidth]{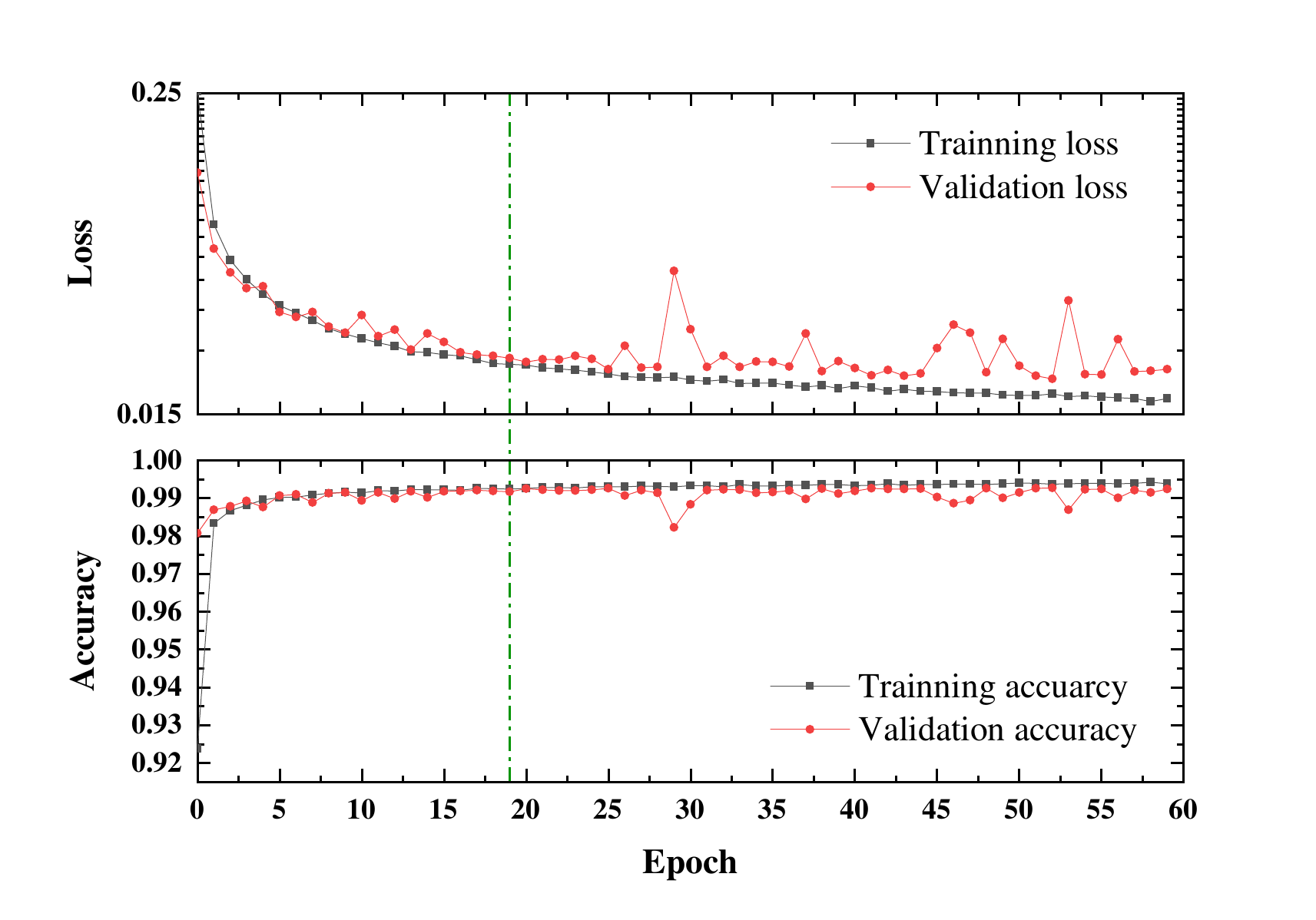}
  \label{fig:trainingProcess-InceptionTime}
}
\hfill
\subfigure[Training loss and accuracy of the EfficientNetV2 model.]{
  \includegraphics[width=0.6\textwidth]{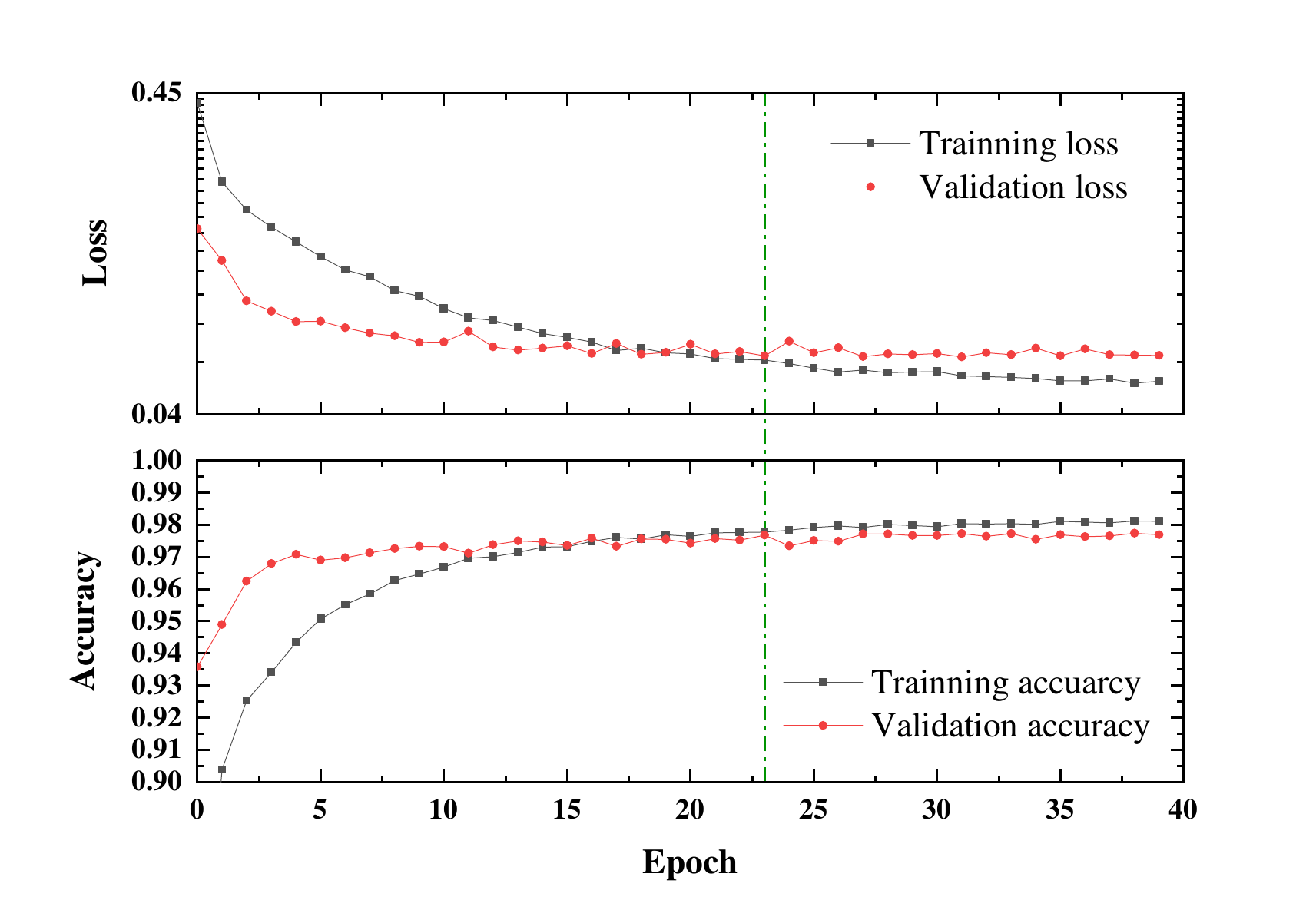}
  \label{fig:trainningProcess-EfficientNetV2}
}
\caption{Loss and accuracy during training.}
\label{fig:trainningProces}
\end{figure}

After several rounds of training and optimization, the model’s hyper-parameters were determined, listed in Tab.\,\ref{tab:optimal_hp}.
The InceptionTime model’s changes in loss and accuracy during training are shown in Fig.~\ref{fig:trainingProcess-InceptionTime}. Here, loss refers to the use of the cross-entropy loss function.
The model’s loss decreases steadily, and the accuracy consistently exceeds 99\%.
Up to the 20th epoch, the model consistently performed on training and validation sets.
However, post the 20th epoch, the divergence between training and testing losses indicated the onset of overfitting.
We chose the 19th epoch’s results as our final model, achieving 99.3\% accuracy on the training set and 99.2\% on the validation set.
A similar procedure was applied to the EfficientNetV2 model, with its training results presented in Fig.~\ref{fig:trainningProcess-EfficientNetV2}. The optimized accuracies are listed in Tab.\,\ref{tab:training results}, together with the results from the InceptionTime model for comparison. The slightly worse performance of EfficientNetV2 model can be attributed to several factors. Firstly, InceptionTime excels in time series classification tasks due to its purpose-built multi-scale feature extraction structure, which enhances its ability to capture crucial information during time series processing. On the other hand, the EfficientNetV2 model is large in scale compared to InceptionTime, leading to potential limitations in hardware resources that may impact hyperparameter optimization and training effectiveness.

\begin{table}[!htb]
\centering
\caption{Results of optimal hyper-parameter tuning.}
\label{tab:optimal_hp}
\begin{tabular}{lccc}
\toprule
Method & Batch size & Learning rate & Optimizer \\
\midrule
InceptionTime & 256 & $5 \times 10^{-6}$ & Adam \\
EfficientNetV2 & 8 & $2 \times 10^{-5}$ & SGD \\
\bottomrule
\end{tabular}
\end{table}

\begin{table}[!htb]
\caption{The accuracy of the training set and the verification set obtained by using different models.}
\label{tab:training results}
\centering
\begin{tabular}{lccc}
\toprule
Method & Optimal epoch & Training accuracy & Validation accuracy \\
\midrule
InceptionTime & 19th & 99.3\% & 99.2\% \\
EfficientNetV2 & 23th & 97.7\% & 97.7\% \\
\bottomrule
\end{tabular}
\end{table}

\section{Results and Discussion}

\begin{figure}
\centering
\subfigure[PSD distribution versus reconstructed energy with color indicating the evaluated InceptionTime CNN score. The vertical dashed line represents the 200\,keVee energy threshold.]{
  \hspace{0.8cm}\includegraphics[width=0.45\textwidth]{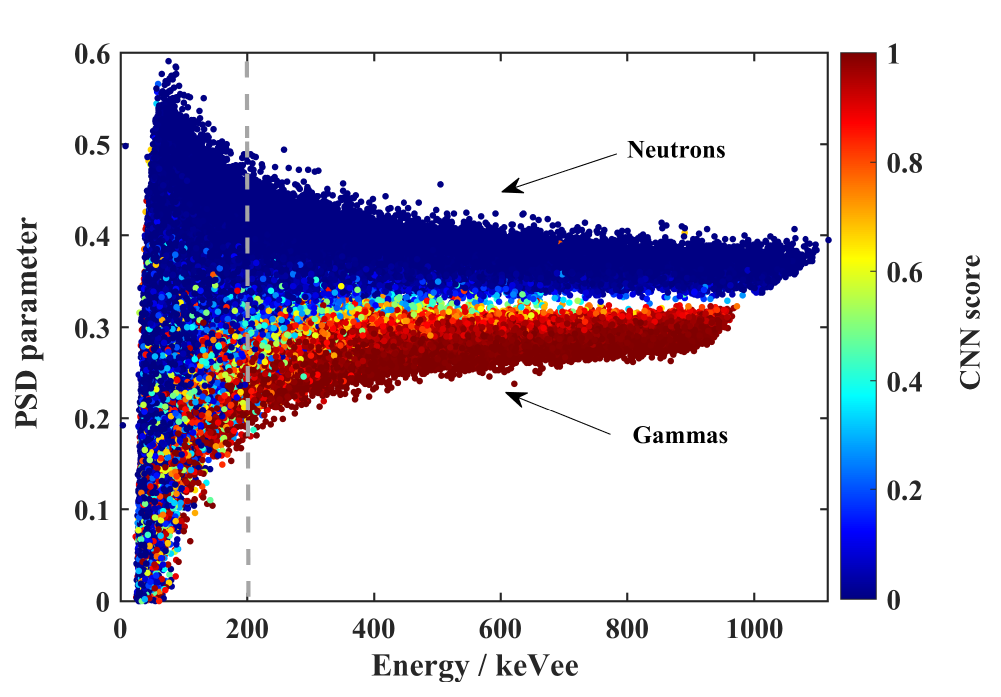}
  \label{fig:sub1}
}
\hfill
\subfigure[Distribution of PSD parameters using CCM method with an energy threshold of 200\,keVee.]{
  \includegraphics[width=0.45\textwidth]{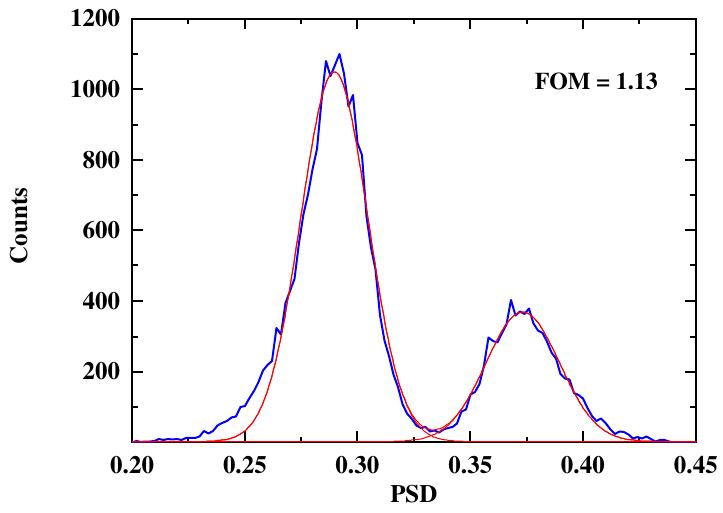}
  \label{fig:sub2}
}
\hfill
\subfigure[Distribution of InceptionTime CNN score with AmBe calibration data. The top panel represents the data with a 200\,keVee energy threshold, while the bottom panel includes the entire energy range.]{
  \hspace{0.5cm}\includegraphics[width=0.45\textwidth]{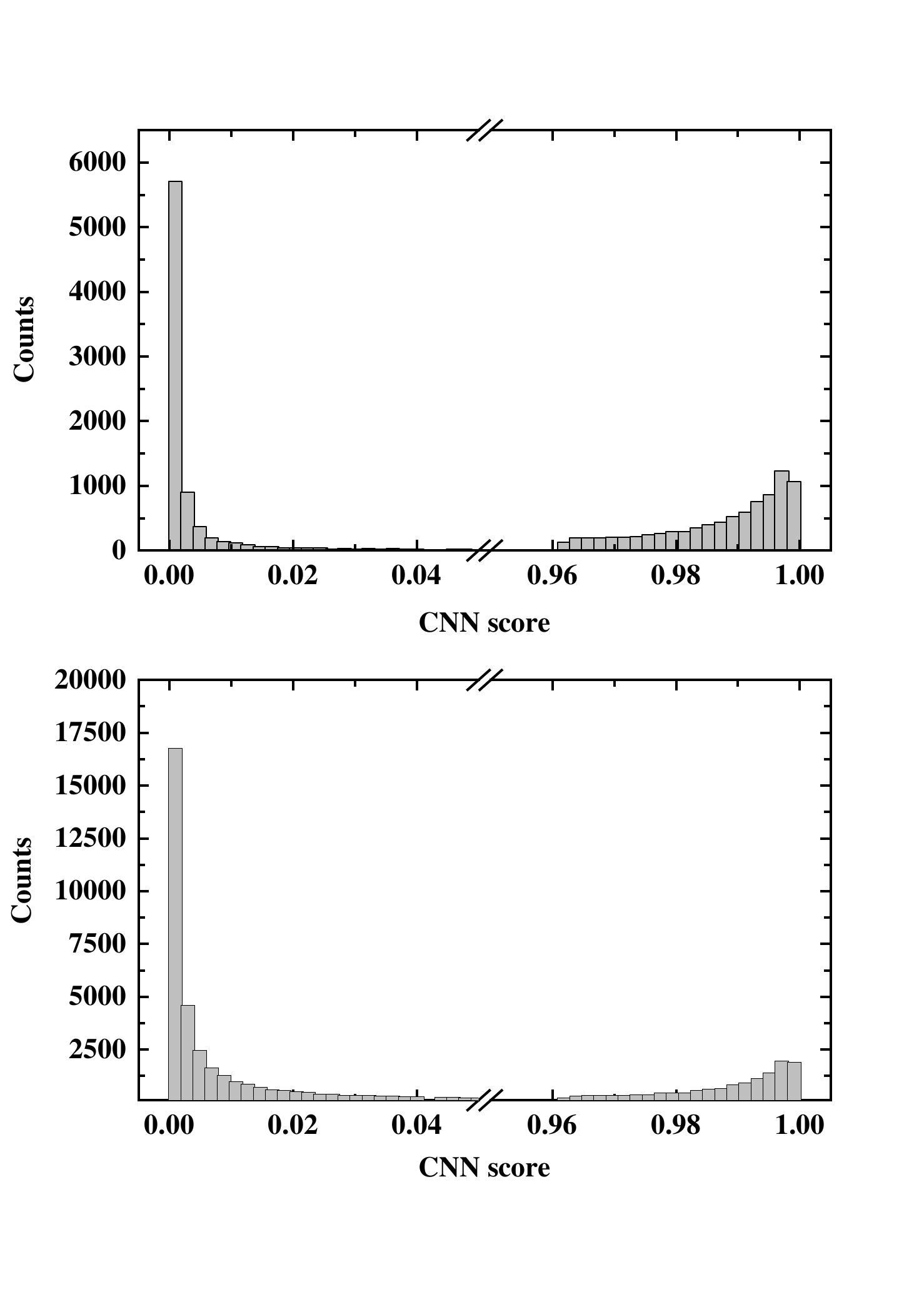}
  \label{fig:sub3}
}

\caption{Distribution of PSD using traditional and InceptionTime methods.}
\label{fig:PSD-CNN}
\end{figure}

Based on the CNN methods mentioned above, we achieved the discrimination result as shown in Fig.~\ref{fig:sub1}.
The CCM offers excellent discrimination capability above 200 keVee, although its effectiveness declines at lower energies. The colour indicates the InceptionTime score for $n$/$\gamma$ discrimination. The closer to the 1(0) score, the more inclined to be an $n$($\gamma$) particle. The distribution of CCM's PSD and CNN score is shown in Fig.~\ref{fig:sub2} and \ref{fig:sub3}, respectively. In terms of the CNN score, most events cluster near 0 or 1, with only a few in the middle range demonstrating a better discrimination capability compared with the CCM method.

\begin{figure}
\centering
\includegraphics[width=0.5\textwidth]{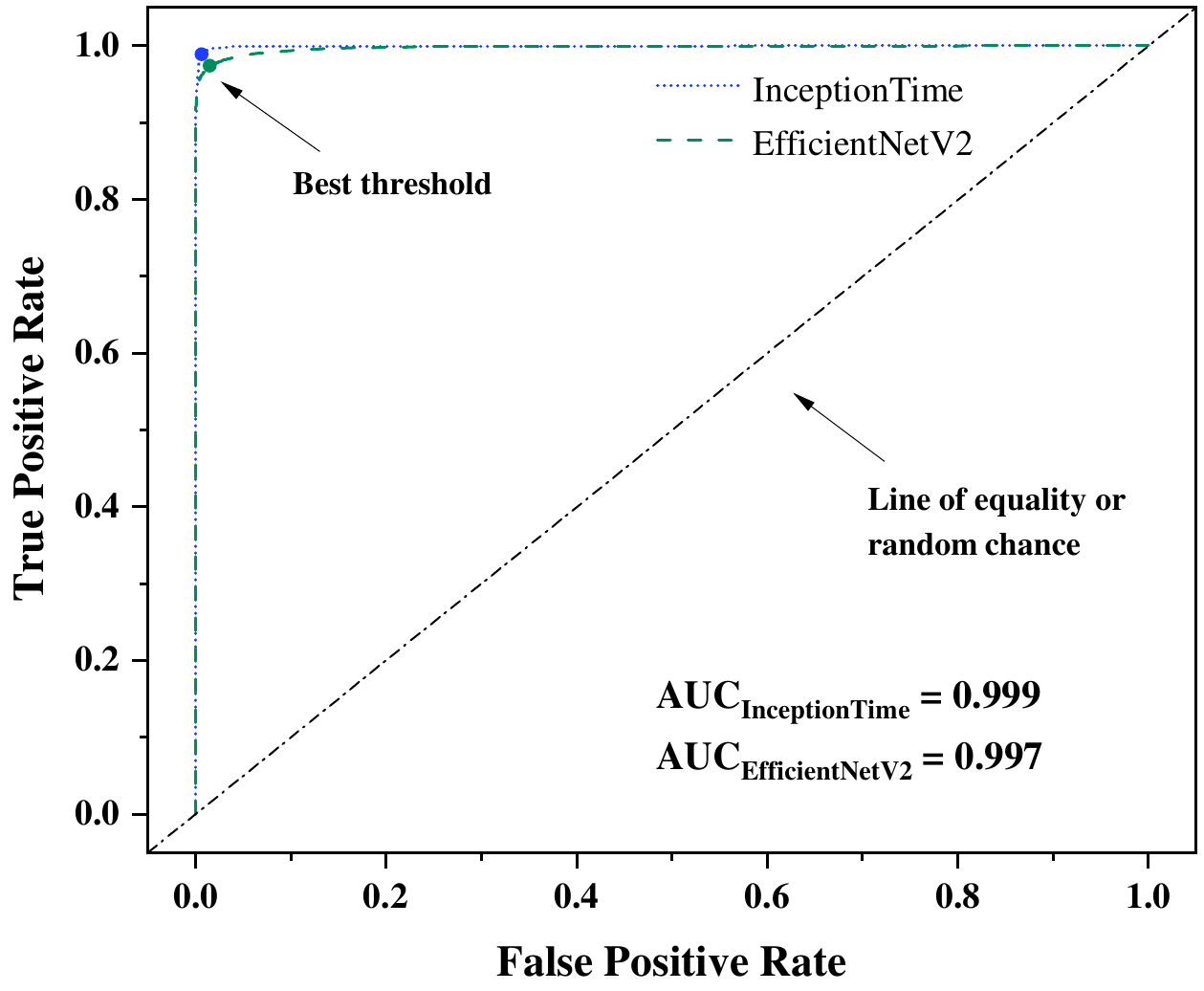}
\caption{ROC curve of CNN method. The blue line represents InceptionTime, and the green line represents EfficientNetV2. Their optimal thresholds are 0.434 and 0.505, respectively.}
\label{fig:ROC}
\end{figure}

Using Eq.\eqref{eq:FoM}, the FoM values calculated are 1.13 for CCM. The InceptionTime CNN method notably provides better separation between neutrons and gamma rays. Without the 200\,keVee energy threshold, the CNN score distribution still maintains good separation. Moreover, to comprehensively assess the model's performance under various discrimination thresholds and identify the optimal one, the Receiver Operating Characteristic (ROC) curve was employed for evaluation. The ROC curves for the two CNN methods were compared, as shown in Fig.~\ref{fig:ROC}. The optimal threshold values were 0.434 for InceptionTime and 0.505 for EfficientNetV2. InceptionTime performed slightly better than EfficientNetV2, with an AUC of 0.999, indicating near-perfect performance.

To compare the CNN method with CCM and assess their performance across different energy regions, we used a gamma source $^{137}$Cs as the test set and evaluated their performance using the accuracy index. Accuracy refers to the recall rate of gamma rays (RG), defined as the proportion of actual gamma ray samples correctly predicted as gamma rays, expressed as
\begin{equation}
\label{eq:accuracy}
\begin{aligned}
RG=\frac{TG}{TG + FN}
\end{aligned}
\end{equation}
TG (true positive) represents the number of correctly classified gamma rays, while FN (false negative) represents the number of gamma rays misclassified as neutrons.

\begin{figure}
    \centering
    \includegraphics[width=0.7\hsize]{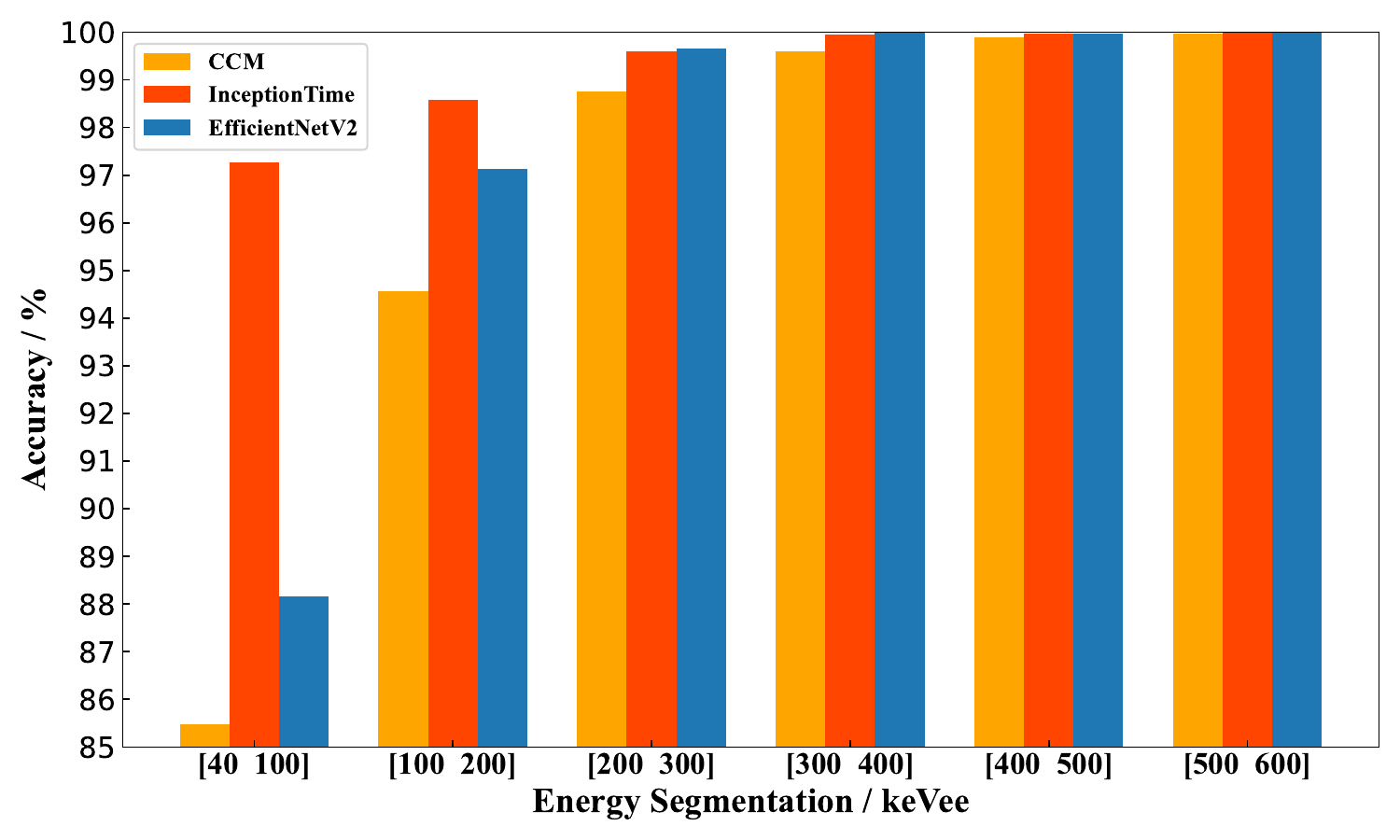}
    \caption{Accuracy rates as a function of energy using CCM and CNN methods. The CNN method shows good performance in low-energy region (\textless\,200\,keVee) for both InceptionTime and EfficientNetV2 models.}
    \label{fig:accuracy}
\end{figure}

Fig.~\ref{fig:accuracy} shows the accuracy rates from both CCM and CNN methods with different energy segmentation. The CCM method used 0.333 as the threshold to distinguish between neutrons and gamma rays, which can be obtained from Fig.~\ref{fig:sub2}. Whereas the CNN method determines the optimal threshold using the ROC curve. This figure highlights that deep learning models discriminate in low-energy segments (below 200\,keVee), significantly outperforming traditional CCM methods. Although all models show similar performance at higher energies (400$\sim$600\,keVee), CNN models demonstrate notable accuracy improvements at around 100\,keVee compared to CCM. Specifically, the InceptionTime model achieves 97.3\% (98.6\%) accuracy below 100\,keVee (100$\sim$200\,keVee), making a 13.8\% (4.25\%) improvement over traditional methods.  This improvement can be attributed to the unique network architecture of InceptionTimeNet, which uses a multiscale convolutional block structure to effectively capture features from the input time series at various temporal scales, identifying complex patterns. This structure excels at detecting faint signals and subtle patterns in low-energy events, a capability that other models may lack. Moreover, InceptionTimeNet's robustness to fluctuations and noise significantly enhances its performance in processing low-energy event signals. Consequently, InceptionTimeNet's superior performance at low-energy levels highlights its advantages in managing data with low signal-to-noise ratios. These findings emphasize the capability of deep learning models to accurately detect low-energy physical events, surpassing traditional methods where they fall short.

\section{Conclusion}
This study is based on a system using a 1-inch EJ-276 plastic scintillator coupled with a SiPM readout. The research compared and analyzed the PSD performance of the CCM and CNN methods. In the CNN methodologies, two distinct inputs and models were used: one-dimensional time series for the InceptionTime model and two-dimensional matrix images for the EfficientNetV2 model. The transformation from one-dimensional to two-dimensional was achieved using RP. The final outcomes show that while the CCM provides good discrimination in the mid to high-energy range, it underperforms in the low-energy region. Conversely, CNN methods display robust discrimination across the entire energy spectrum, particularly the InceptionTime method, which significantly outshines other methods in low-energy scenarios. The superiority of the CNN method ensures accurate threat detection, particularly in homeland security applications. Additionally, improved particle discrimination in the low-energy region aids in detecting dark matter, enhancing veto efficiency, and further suppressing background noise.

\acknowledgments

This work was supported by grants from the China Postdoctoral Science Foundation (2023M744093), the National Natural Science Foundation of China (No.12075330), the National Key R\&D Program of China (No.2022YFC2402300), and the Fundamental Research Funds for the Central Universities, Sun Yat-sen University (No.22lglj11).


\end{document}